\documentclass[aps,prc,preprint,superscriptaddress,showpacs]{revtex4}
\usepackage[dvipdfmx]{graphicx}
\usepackage{amsmath}
\usepackage{amssymb}
\usepackage{epstopdf}
\renewcommand{\theequation}
{{\rm \thesection.\arabic{equation}}}
\newcommand{\bm}[1]{\mbox{\boldmath $ #1 $}}
\def\be{\begin{equation}}
\def\ee{\end{equation}}
\def\bea{\begin{eqnarray}}
\def\eea{\end{eqnarray}}

\def\eps{\varepsilon}
\def\eps0{\varepsilon_0}

\begin{document}
\title{Recursion Method for Deriving  Energy-Independent Effective Interaction}
\author{Kenji Suzuki}
\email[]{suz93@mocha.ocn.ne.jp}
\affiliation{Senior Academy, Kyushu Institute of Technology, 
             Kitakyushu 804-8550, Japan}
\author{Hiroo Kumagai}
\email[]{kumagai@fit.ac.jp}
\affiliation{Faculty of Information Engineering, Fukuoka Institute of Technology, 
             Fukuoka 811-0295, Japan}
\author{Ryoji Okamoto}
\email[]{okamoto.ryoji.munakata@gmail.com}
\affiliation{Senior Academy, Kyushu Institute of Technology, 
             Kitakyushu 804-8550, Japan}
\author{Masayuki Matsuzaki}
\email[]{matsuza@fukuoka-edu.ac.jp}
\affiliation{Department of Physics, Fukuoka University of Education, 
             Munakata, Fukuoka 811-4192, Japan}     
\date{\today}
\begin{abstract}
    The effective-interaction theory has been one of the useful and 
practical methods for solving nuclear many-body problems based on the 
shell model.   
Various approaches have been proposed which are constructed 
in terms of the so-called $\widehat{Q}$ box and its energy derivatives introduced 
by Kuo {\it et al}.  
In order to find out a method of calculating them we make
decomposition of a full Hilbert space into subspaces (the Krylov subspaces) and
transform a Hamiltonian to a block-tridiagonal form.  
This transformation brings about much simplification of the calculation of the 
$\widehat{Q}$ box.  
In the previous work a recursion method has been derived for calculating the $\widehat{Q}$ box 
analytically on the basis of 
such transformation of the Hamiltonian.   
In the present study, by extending the 
recursion method for the $\widehat{Q}$ box, we derive another recursion relation 
to calculate the derivatives of the $\widehat{Q}$ box of arbitrary order. 
With the $\widehat{Q}$ box and its derivatives thus determined we apply them to the
calculation of the $E$-independent effective interaction given in the so-called 
Lee-Suzuki (LS) method for a system with a degenerate unperturbed energy.  
We show that the recursion method can also be applied to the generalized LS 
scheme for a system with non-degenerate unperturbed energies.  
If the Hilbert 
space is taken to be sufficiently large, the theory provides an exact way of 
calculating the $\widehat{Q}$ box and its derivatives.  
This approach enables us to perform recursive calculations for the effective interaction 
to arbitrary order for both systems with degenerate and non-degenerate unperturbed energies.
\end{abstract}
\maketitle
%
%
\section{Introduction}
   In nuclear many-body physics based on the shell model it is usually necessary to introduce a small-dimensional
model space and recast the full shell-model calculation in the form 
of a model-space problem with an effective interaction.
The effective-interaction theory has been developed as regards formal theories and 
their actual applications for more than forty years.  
The progress in this field of physics has been reviewed in the recent 
articles given by Coraggio {\it et al}.~\cite{CCGIK09, CCGIK12}.

    The first attempt to construct an effective interaction was made by
Bloch and Horowitz~\cite{BH58} and Feshbach~\cite{Fes62}.   
They derived an energy ($E$)-dependent form of the effective interaction.   
The perturbation expansion of this $E$-dependent form has also been studied and 
called the Brillouin-Wigner expansion~\cite{LM86, WH12}.   
This $E$-dependent effective interaction has been
recognized as an original and basic form, but the $E$-dependence is not always
desirable because the effective interaction contains an uncertain variable $E$
which is the eigenvalue of a given Hamiltonian to be solved.

    Elimination of the $E$-dependence was made, on the basis of the perturbation
expansion, in two ways:  
One is the time-independent approach by Brandow~\cite{Bra67},
and the other is the time-dependent one by Kuo, Lee and Ratcliff~\cite{KLR71}.  
They have proved that the $E$-dependence could be removed by introducing special
type of diagrams, the folded diagrams~\cite{Mor63, OOR63}.  
It has been clarified that the $E$-independent effective interaction consists of two series 
of diagrams, namely, the usual linked non-folded diagrams and the folded ones.

    A next problem was how to sum up the perturbation series including the
folded diagrams.   Kuo and others~[8] defined a vertex function called the
$\widehat{Q}$ box as the sum of linked and non-folded diagrams.   They have proved that
the folding operation in diagrams corresponds to the differentiation
with respect to $E$ which is the diagram starting energy (unperturbed energy).
Resultantly it has been established that the $E$-independent effective 
interaction can be represented as a series expansion in terms of only the
$\widehat{Q}$ box and its derivatives.

    The folded-diagram theory based on the $\widehat{Q}$ box has succeeded 
in deriving the $E$-independent form of the effective interaction, 
but it still consists of a series of terms to infinite order.  
A problem is then how to sum up the series.  The iteration
methods have been introduced to sum up partially the folded diagrams.   Two
approaches have been known.   One is of Krenciglowa and Kuo (KK)~\cite{KK74}, who 
introduced a partial summation method for the folded diagrams and
derived a self-consistent equation for  the effective interaction and 
eigenvalues of a given Hamiltonian.  
Recently, defining a new vertex function, 
namely the $\widehat{Z}$ box, an extended KK approach~\cite{SOKF11} has been proposed and
applied to some actual cases~\cite{CCGIK12}.   As is well known, the KK iterative equation
reproduces only the eigenvalues of the eigenstates which have largest 
overlaps with the model space although this property of convergence has not yet been proved
theoretically and an exceptional case has been reported by Takayanagi~\cite{Tak11}.  
 On the other hand, the $\widehat{Z}$-box method is a 
state-independent theory, that is, any set of the eigenvalues of the 
Hamiltonian can be reproduced regardless of  properties of the eigenstates.
The $\widehat{Z}$ box is an $E$-dependent operator given with the $\widehat{Q}$ box and its first 
derivative.  The $\widehat{Z}$ box becomes the effective interaction when it is calculated
at the energy variable $E$ which coincides with one of the true eigenvalues of
the Hamiltonian.  In this sense the $\widehat{Z}$ box method is not an $E$-independent theory.
This situation of the $\widehat{Z}$ box method is the same as the KK approach.

    The other is of Lee and Suzuki (LS)~\cite{LS80, SL80}, who derived an effective 
interaction by means of the similarlity-transformation theory in the
eigenvalue problem.   They set up a general equation for determining 
the effective interaction, which is often referred to as the decoupling 
equation.  An equivalent equation was also given by Okubo~\cite{Oku54} in a
different way.  The decoupling equation was solved for a system 
with a degenerate unperturbed energy and a recursive solution was derived.  
The LS solution is completely $E$-independent, that is, it does not 
contain any uncertain energies.    The solution has been given 
in terms of the $\widehat{Q}$ box and its derivatives at a fixed energy variable 
$E$ being the unperturbed energy.

    The LS method has been generalized 
to the case with non-degenerate unperturbed energies by Suzuki, Okamoto, 
Ellis and Kuo~\cite{SOEK94}, which we refer to as the generalized Lee-Suzuki (GLS) method. 
It has been shown that the GLS scheme yields a new recursive method in which $d$ initial
values for $E$'s, the unperturbed energies, can be introduced as input 
parameters, where $d$ is the dimension of the model space.   This situation
of the GLS method is significantly different from the LS case in which only
one input parameter can be taken as an initial energy.
   
   In view of the effective-interaction theories given to date, we 
focus our attention on the LS and GLS methods in order to construct an
$E$-independent effective interaction.  
  Both of the solutions are derived 
by solving the decoupling equation and represented in terms of the $\widehat{Q}$ box 
and its derivatives.  
  The problem is now how to calculate the $\widehat{Q}$ box and 
how to differentiate it.  
  As for the $\widehat{Q}$ box, the most widely applied 
method has been the perturbation theory.   
  The diagrams have been taken into account up to third order~\cite{CCGIK09, CCGIK12, HKO95} 
and partially to fourth order~\cite{BKO70}.
  The evaluation of diagrams of more-than-fourth order has been considered to be
prohibitive due to limitation of computer facilities~\cite{CCGIK12}.   
  The convergence of order-by-order calculations has been investigated by many authors.   
  The present status of the perturbative calculations has been reviewed 
in Refs.\cite{CCGIK09, CCGIK12}.
  Some of the numerical calculations for two-valence-particle systems, such as
$^{18}$O and $^{134}$Sn, have shown that the second-order contribution is dominant and
higher-order terms are less important, if the single-particle energies in 
the energy denominators are replaced with the experimental ones.   
  In the previous study by the authors~\cite{SKMO13} a non-perturbative approach to the
calculation of the $\widehat{Q}$ box has been presented.   
They have shown that the $\widehat{Q}$ box
can be expressed as second-order perturbation terms if a proper
renormalized energy denominator is given.   They have suggested
that there is a possibility that the energy denominator can be given
approximately in terms of the exact single-particle energies, that is, the
experimental ones.

    The remaining task in constructing the $E$-independent effective interaction,
such as in the LS method, is to calculate the energy derivatives of the $\widehat{Q}$ box. 
  Up to date the method of numerical differentiation has been employed~\cite{HKO95}.  
  In general, higher-order calculations in the LS recursion method require 
higher-order differentiations of the $\widehat{Q}$ box, and therefore we need to calculate
the $\widehat{Q}$ box at many points of the energy variable $E$.   
  This situation in the LS scheme might loose the advantage of the $E$ independence.  
A main purpose of the present study is to derive an analytical and
non-perturbative method of calculating the energy derivatives by following the
previous work in which a recursive equation has been given for the 
non-perturbative $\widehat{Q}$ box.   This approach enables us to calculate the derivatives
of the $\widehat{Q}$ box of arbitrary order and to obtain resultantly the LS recursive
solution of much higher order.   

    The recursion method for the LS solution can
be shown to be extended to a general system with non-degenerate unperturbed 
energies.   
For the GLS case we shall prove that the recursive solution can also
 be given in terms of only the $\widehat{Q}$ box and its derivatives at energy
variables $E$'s being unperturbed energies.  
In this case of the GLS method
$d$ energy variables of $E$'s can be introduced as input parameters.   
For this reason there are
wide possibilities of selecting the initial energies and, therefore,  faster 
convergence in the recursive calculation can be expected  than that of the LS method.  
 We finally verify this prediction for convergence by performing a numerical
calculation with a simple model Hamiltonian.

    The organization of the present article is as follows:  
In Sec.II, we briefly outline the standard method for constructing the effective
interaction.   The recursion relations for the LS and GLS solutions are derived 
for the systems with degenerate and non-degenerate energies, respectively. 
We show that both of the LS and GLS solutions can be given with the $\widehat{Q}$ box 
and its derivatives.
   In Sec.III, we derive recursion methods for calculating the $\widehat{Q}$ box 
and its derivatives in an analytic and non-perturbative way.
   In Sec.IV, a model calculation is made to obtain some assessments of the present
approach by introducing a simple model Hamiltonian.
   A summary and some concluding remarks are given in the last section.  
%

%
\section{Derivation of Energy-independent effective interaction}
%
%
\subsection{Basic equations for the effective interaction}
We start with a Hamiltonian $H$ defined in a Hilbert space.
We divide the space into two subspaces, namely, the model space and the complementary 
space which are referred to as the $P$ and $Q$ spaces, respectively.
If all the eigenvalues of an operator $H_{\rm eff}$ defined in the $P$ space coincide 
with those of the original Hamiltonian $H$, we call $H_{\rm eff}$ the effective Hamiltonian.

There have been various ways of deriving $H_{\rm eff}$. We here adopt the standard method
given in Ref.\cite{SL80}.
We consider an operator $\omega$ which maps states in the $P$ space and those in the $Q$ space
 to each other. 
The operator $\omega$ has the following properties:
\begin{align}
\label{eq:omega-definition}
 \omega &= Q\omega P, \\
 \label{eq:omega-property}
 \omega^{n} &= 0 \hspace{10mm} (n\ge 2)
\end{align}
 and 
\begin{align}
\label{eq:e-omega-expansion}
 e^{\omega} &= 1 + \omega.
\end{align}
 
With the operator $\omega$ we define a similarity transformation of $H$ as
\begin{align}
\label{eq:H-transformation}
 \widetilde{H} &= {\rm e}^{-\omega}H {\rm e}^{\omega} \nonumber\\
                      &= (1-\omega)H(1+\omega).
\end{align}
The condition that $P\widetilde{H}P$ be an effective Hamiltonian is that $\widetilde{H}$ 
should be decoupled
between the $P$ and $Q$ spaces as
\begin{align}
\label{eq:decoupling-eq1}
 Q \widetilde{H} P &=  0.
\end{align}
The above condition leads to an equation for $\omega$ written as 
\begin{align}
\label{eq:decoupling-eq2}
 QHP+QHQ\omega-\omega PHP-\omega PHQ\omega&= 0,
\end{align}
which has been called the decoupling equation and was also derived by Okubo~[16].
Once a solution $\omega$ to Eq.(\ref{eq:decoupling-eq2}) is given, the effective Hamiltonian $H_{\rm eff}$ is written
 as
\begin{align}
\label{eq:definition-Veff}
    H_{\rm eff} &= P \widetilde{H} P \nonumber\\
                      &= PHP + PHQ\omega .
\end{align}
Dividing $PHP$ into the unperturbed part $PH_0P$ and the interaction $PVP$, we write $PHP$ 
as  
\begin{align}
\label{eq:divide-PHP}
    PHP= PH_{0}P + PVP.
\end{align}
The effective interaction $R$ is introduced through 
\begin{align}
\label{eq:definition-Heff-R}
    H_{\rm eff} &= PH_0P + R
\end{align}
and therefore
\begin{align}
\label{eq:definition-R}
    R=PVP + PHQ\omega.
\end{align}
From the above formulation we see that the construction of the effective interaction $R$
reduces to solving the  decoupling equation (\ref{eq:decoupling-eq2}) for $\omega$.

    In a recent work by Takayanagi~\cite{Tak13} it has been proved that the 
decoupling equation provides a necessary and sufficient condition of 
determining the effective interaction which reproduces any set of $d$ eigenvalues
of the Hamiltonian $H$, where $d$ is the dimension of the $P$ space. 
This rigorous proof demonstrates that any of the effective interactions such as those 
in  the KK, LS, GLS and Andreozzi's methods~\cite{And87}, should be derived 
as the solutions of the decoupling equation.   
\subsection{Effective interaction with degenerate unperturbed energy}
We  consider a system with a degenerate unperturbed energy. We write the Hamiltonian 
of the system as
\begin{align}
\label{eq:degenerate-Hamiltonian}
H = E_0P + PVP + PHQ + QHP + QHQ,
\end{align}
where $E_0$ is the unperturbed energy.
An $E$-independent effective interaction for the degenerate system has been derived in 
Ref.\cite{SL80}. We briefly outline the derivation.
We define three operators as functions of an energy variable $E$ as
\begin{align}
\label{eq:eE}
       e(E) &= Q(E-H)Q, \\
\label{eq:Q-box}
 \widehat{Q} (E)&= PVP +PHQ
                           \frac{1}{E-QHQ}QHP,
\end{align} 
and
\begin{align}
\label{eq:another-definition1-Q-box}
 \widehat{Q}_k (E) &= (-1)^k PHQ\frac{1}{(E-QHQ)^{k+1}}QHP \cr
                              &= \frac{1}{k!}\frac{d^k}{dE^k}\widehat{Q}(E).
\end{align}
 The operator $\widehat{Q}(E)$ is called the $\widehat Q$ box according to Kuo and 
his collaborators \cite{KLR71}.

Using the decoupling equation (\ref{eq:decoupling-eq2}) and Eq.(\ref{eq:definition-R}) 
for the effective interaction $R$, we have a formal solution for $\omega$ given by
\begin{align}
\label{eq:omega-formal-solution}
\omega = \frac{1}{e(E_0)}QHP - \frac{1}{e(E_0)}\omega R,
\end{align}
where
\begin{align}
\label{eq:definition-e(E0)}
e(E_0) = E_0 -QHQ.
\end{align}

The $E$-independent solution for $R$ is obtained from the following recursion relations
\begin{align}
\label{eq:omega-recursion-formula-1}
\omega_n &= \frac{1}{e(E_0)}QHP - \frac{1}{e(E_0)}\omega_{n-1} R_n
\end{align}
and
\begin{align}
\label{eq:definition-Rn}
R_n &= PVP + PHQ\omega_n\nonumber\\
    &=\left[ P+PHQ\frac{1}{e(E_{0})}\omega_{n-1}\right]^{-1} \widehat{Q}(E_{0}).
\end{align}
Starting with $\omega_0 =0$, we have from  Eq.(\ref{eq:omega-recursion-formula-1})
\begin{align}
\label{eq:omega-recursion-formula-2}
\omega_n = \frac{1}{e}QHP - \frac{1}{e^2}QHP R_n +\cdots
                    +(-)^{n+1}\frac{1}{e^n}QHP R_2 R_3 \cdots R_n,
\end{align}
where we have used for simplicity
\begin{align}
\label{eq:another-expression-e(E0)}
e = e(E_0).
\end{align}
Substituting $\omega_n$ in Eq.(\ref{eq:omega-recursion-formula-2}) into 
Eq.(\ref{eq:omega-recursion-formula-1}), the recursive solution for $R_n$ is solved as 
\begin{align}
\label{eq:recursive-solution-Rn}
R_n =\left[1 - \widehat{Q}_1 -  \widehat{Q}_2R_{n-1} - \cdots  
- \widehat{Q}_{n-1}R_2R_3 \cdots R_{n-1} \right]^{-1}\cdot  \widehat{Q} \hspace{1cm} n\geq 3,
\end{align}
where we have used the abbreviations as 
\begin{align}
\label{eq:abbreriation-Q}
\widehat{Q} &=  \widehat{Q}(E_0)\,
\end{align}
and
\begin{align}
\label{eq:abbreriation-Qk}
\widehat{Q}_k &=  \widehat{Q}_k(E_0).
 \end{align}

In Eq.(\ref{eq:recursive-solution-Rn}), the initial values for $n=1, 2$ are
\begin{align}
\label{eq:R1}
R_1 &= \widehat{Q}(E_0),\\
\label{eq:R2}
R_2 &= \left[1 - \widehat{Q}_1(E_0) \right]^{-1}\cdot \widehat{Q}(E_0). 
 \end{align}
The recursion relation Eq.(\ref{eq:recursive-solution-Rn}) determines the sequence 
$\{ R_n,\,\,\, n=1,2, \cdots\}$. 
If $R_n$ converges when $n$ tends to infinity, the effective interaction $R$ is given by 
\begin{align}
\label{eq:R-infinity}
R & =  R_{\infty} \cr
   & =\lim_{n\rightarrow \infty}R_n.
 \end{align}
This $E$-independent effective interaction has been known as Lee-Suzuki's  (L-S)
solution~\cite{LS80, SL80}. 
The convergence condition for this recursion method has been discussed in Ref.\cite{SL80}.
It has been known that the effective interaction $R$ reproduces the eigenvalues of $H$ 
which are the nearest to the unperturbed energy $E_{0}$.

In order to obtain the sequence $\{ R_{n}, n=1,2,\cdots\}$ we have to calculate the operators 
$\widehat{Q}(E_{0})$ and $\{ \widehat{Q}_{k}(E_{0}), k=1,2,\cdots\}$. Therefore,
the procedure of obtaining the $E$-independent effective interaction for a
degenerate system can be reduced to calculating 
$\widehat{Q}(E_{0})$ and $\{ \widehat{Q}_{k}(E_{0}), k=1,2,\cdots\}$. 
In the next section we show that these operators can be calculated analytically
by means of recursion methods. 
\subsection{Effective interaction with non-degenerate unperturbed energies}
In actual problems there are many cases in which the unperturbed energies are not
degenerate. In this case  the Hamiltonian is given generally as
\begin{eqnarray}
\label{eq:VII.1}
 H&=&H_{0}+PVP+PHQ+QHP+QHQ
\end{eqnarray}
with
\begin{eqnarray}
\label{eq:VII.2}
H_{0}&=&\sum_{\alpha} \varepsilon_{\alpha}|\phi_{\alpha}^{(0)}\rangle 
                               \langle\tilde{\phi}_{\alpha}^{(0)} |,
\end{eqnarray}
where $\varepsilon_{\alpha}$ and $|\phi_{\alpha}^{(0)}\rangle$ are the unperturbed 
energy and state, respectively. 
The $\langle\tilde\phi_{\alpha}^{(0)}|$ is the biorthogonal state of $|\phi_{\alpha}^{(0)}\rangle$.
We define a projection operator
\begin{eqnarray}
\label{eq:VII.3}
P_{\alpha}&=&|\phi_{\alpha}^{(0)}\rangle 
                           \langle\tilde{\phi}_{\alpha}^{(0)} |,
\end{eqnarray}
then $H_{0}$ is written as 
\begin{eqnarray}
\label{eq:VII.4}
H_{0}&=&\sum_{\alpha} \varepsilon_{\alpha}P_{\alpha}.
\end{eqnarray}

In the non-degenerate case the formal solution for $\omega$ can be derived from
Eqs.(\ref{eq:decoupling-eq2}) and (\ref{eq:definition-R}) as
\begin{eqnarray}
\label{eq:VII.5}
\omega &=& \sum_{\alpha} 
 \left[  \frac{1}{e(\epsilon_{\alpha})}QHP_{\alpha}
        -\frac{1}{e(\epsilon_{\alpha})}\omega RP_{\alpha}
\right],
\end{eqnarray}
where
\begin{eqnarray}
\label{eq:VII.6}
e(\epsilon_{\alpha})&=&\epsilon_{\alpha}-QHQ.
\end{eqnarray}
Inserting $\omega$ in Eq.(\ref{eq:VII.5}) into Eq.(\ref{eq:definition-R}) we have an equation 
for $R$ given by
\begin{eqnarray}
\label{eq:VII.7}
R &=& PVP+\sum_{\alpha} PHQ\frac{1}
                                                 {e (\epsilon_{\alpha})}QHP_{\alpha}
                -\sum_{\alpha} PHQ\frac{1}
                                                 {e (\epsilon_{\alpha})}\omega RP_{\alpha}
\end{eqnarray}
We then have a formal solution for $R$ as
\begin{eqnarray}
\label{eq:VII.8}
R &=& \sum_{\alpha} \left[ 
                                     P+PHQ\frac{1}
                                                      {e (\epsilon_{\alpha})}\omega
                            \right]^{-1}\widehat{Q}(\epsilon_{\alpha})P_{\alpha},
\end{eqnarray}
where $\widehat{Q}(\epsilon_{\alpha})$ is the $\widehat{Q}$ box in Eq.(\ref{eq:Q-box}) with 
the energy variable 
$E=\epsilon_{\alpha}$.
With Eqs.(\ref{eq:VII.5}) and (\ref{eq:VII.8}) we set up recurrence equations for 
$\omega$ and $R$ as
\begin{eqnarray}
\label{eq:VII.9}
\omega_{n} &=& \sum_{\alpha} 
 \left[  \frac{1}{e(\epsilon_{\alpha})}QHP_{\alpha}
        -\frac{1}{e(\epsilon_{\alpha})}\omega_{n-1} R_{n}P_{\alpha}
\right],
\end{eqnarray}
and
\begin{eqnarray}
\label{eq:VII.10}
R_{n} &=& \sum_{\alpha} \left[ 
                                     P+PHQ\frac{1}
                                                      {e (\epsilon_{\alpha})}\omega_{n-1}
                            \right]^{-1}\widehat{Q}(\epsilon_{\alpha})P_{\alpha}.
\end{eqnarray}
These two recursion relations can be understood as general extensions
of Eqs.(\ref{eq:omega-recursion-formula-1}) and (\ref{eq:definition-Rn}) for a system
with a degenerate unperturbed energy $E_{0}$ to the case 
with non-degenerate unperturbed energies $\{ \varepsilon_{\alpha}, \alpha=1,2,\cdots,d \}$, 
where $d$ is the dimension of the $P$ space. 

We start with the initial value $\omega_{0}=0$. The solutions for $R$ are given by
\begin{eqnarray}
\label{eq:VII.11}
 R_{1}&=&\sum_{\alpha}\widehat{Q}(\epsilon_{\alpha})P_{\alpha}
\end{eqnarray}
and
\begin{eqnarray}
\label{eq:VII.12}
 R_{2}&=&\sum_{\alpha} \left[
                                      P-\sum_{\beta}\widehat{Q}_{1}(\epsilon_{\alpha,}\epsilon_{\beta})P_{\beta}
                               \right]^{-1}
                    \widehat{Q}(\epsilon_{\alpha})P_{\alpha},
\end{eqnarray}
and generally 
\begin{eqnarray}
\label{eq:VII.13}
 R_{n}&=&\sum_{\alpha} X_{n-1, \alpha}\widehat{Q}(\epsilon_{\alpha})P_{\alpha},
\end{eqnarray}
where 
\begin{eqnarray}
\label{eq:VII.14}
 X_{n-1, \alpha}&=&P-\sum_{\beta}\widehat{Q}_{1}(\epsilon_{\alpha},\epsilon_{\beta})P_{\beta}
                   -\sum_{\beta,\gamma}
                        \widehat{Q}_{2}(\epsilon_{\alpha},\epsilon_{\beta},\epsilon_{\gamma})
                       P_{\beta}R_{n-1}P_{\gamma}-\cdots
                       \nonumber\\
              &&-\sum_{\beta,\gamma,\cdots,\lambda,\mu}
                        \widehat{Q}_{n-1}(\epsilon_{\alpha},\epsilon_{\beta},\cdots,\epsilon_{\lambda},
                                                                                                \epsilon_{\mu})
                       P_{\beta}R_{2}P_{\gamma}R_{3}\cdots R_{n-2}P_{\lambda}R_{n-1}P_{\mu},
\end{eqnarray}
and
\begin{eqnarray}
\label{eq:VII.15}
 \widehat{Q}_{m}(\epsilon_{1},\epsilon_{2},\cdots,\epsilon_{m+1})
 &=&(-1)^{m}PHQ\frac{1}
         {e(\epsilon_{1})e(\epsilon_{2})\cdots e(\epsilon_{m})e(\epsilon_{m+1})}QHP.
\end{eqnarray}
The expression of $R$ in Eq.(\ref{eq:VII.13}) is a straightforward extension of $R$ 
in Eq.(\ref{eq:recursive-solution-Rn})
for the degenerate case to the non-degenerate case. This  has already been derived 
by Suzuki, Okamoto, Ellis and Kuo~\cite{SOEK94}, which we call the generalized LS (GLS) solution.

   The problem is now how to calculate 
$\widehat{Q}_{m}(\varepsilon_{1},\varepsilon_{2},\cdots,\varepsilon_{m+1})$ in Eq.(\ref{eq:VII.15})
which we call the multi-energy $\widehat{Q}$ box. We here note that the subscript $m$
 means that the operator $\widehat{Q}_{m}(\varepsilon_{1},\varepsilon_{2},\cdots,\varepsilon_{m+1})$
 contains $m+1$ denominators. Therefore, when all the energies $\{ \varepsilon_{\alpha}, \alpha=1,2,
\cdots, m+1\}$ are the same,  the operator 
$\widehat{Q}_{m}(\varepsilon_{1},\varepsilon_{2},\cdots,\varepsilon_{m+1})$
reduces to
$\widehat{Q}_{m}(\varepsilon_{1})$ with a single energy variable in Eq.(\ref{eq:another-definition1-Q-box}).
In this sense, the multi-energy 
$\widehat{Q}$ box $\widehat{Q}_{m}(\varepsilon_{1},\varepsilon_{2},\cdots,\varepsilon_{m+1})$ 
can also be understood as an extension of the $\widehat{Q}_{m}(\varepsilon_{1})$ 
in Eq.(\ref{eq:another-definition1-Q-box})
with a single energy variable $\varepsilon_{1}$ to the case with multi-energy
variables $(\varepsilon_{1},\varepsilon_{2},\cdots,\varepsilon_{m+1})$.
If the variables $(\varepsilon_{1},\varepsilon_{2},\cdots,\varepsilon_{m+1})$
are all different, the multi-energy $\widehat{Q}_{m}$ box in 
Eq.(\ref{eq:VII.15}) can be expressed as a linear combination of the $\widehat{Q}$ box
as  \cite{SKMO13} 
\begin{align}
\label{eq:Q-box-linear-combi}
\widehat{Q}_m(\varepsilon_{1},\varepsilon_{2},\cdots,\varepsilon_{m+1})
 &=\sum_{k=1}^{m+1}C_{k}(\varepsilon_{1},\varepsilon_{2},\cdots,\varepsilon_{m+1})
   \widehat{Q}(\varepsilon_{k})
\end{align}
with 
\begin{align}
\label{eq:Ck-linear-combi}
C_{k}(\varepsilon_{1},\varepsilon_{2},\cdots,\varepsilon_{m+1})
 &=\prod_{i=1(i\ne k)}^{m+1}\frac{1}{(\varepsilon_{k}-\varepsilon_{i})}. 
\end{align}

We consider a general case that some of the energy variables 
$(\varepsilon_{1},\varepsilon_{2},\cdots,\varepsilon_{m+1})$
are the same. Let $d$ be the dimension of the $P$ space. We assume that 
the unperturbed energies $\{ \varepsilon_{\alpha}, \alpha=1,2,\cdots, d\}$ are all different.
In this case we write the multi-energy $\widehat{Q}_{m}$ box as 
\begin{align}
\label{eq:Q-box-eps-n}
\widehat{Q}_m({\bm \varepsilon}^{(d)},{\bm n}^{(d)})
 &=(-1)^{m}PHQ\frac{1}{e(\varepsilon_{1})^{n_1}e(\varepsilon_{2})^{n_2}\cdots e(\varepsilon_{d})^{n_d}}QHP,
\end{align}
where  ${\bm \varepsilon}^{(d)}$ and ${\bm n}^{(d)}$ are the $d$-dimensional vectors defined as
\begin{align}
\label{eq:bm-varepsilon}
{\bm \varepsilon}^{(d)}&=(\varepsilon_1,\varepsilon_2,\cdots,\varepsilon_d)
\end{align}
and
\begin{align}
\label{eq:bm-n}
{\bm n}^{(d)}&=(n_{1},n_{2},\cdots,n_{d}).
\end{align}
Since $m+1$ means the number of the energy denominators, the numbers $(n_{1},n_{2},\cdots,n_{d})$
should satisfy
\begin{align}
\label{eq:n-sum}
n_1 +n_2+\cdots+n_d &=m+1.
\end{align}

It is proved that  the multi-energy $\widehat{Q}_{m}$ box in Eq.(\ref{eq:Q-box-eps-n}) can be given 
by a linear combination of 
$\{ \widehat{Q}_{k}(\varepsilon_{i}), 1\le k \le m+1, 1\le i\le d\}$ with a single energy
variable. Actually we may write as
\begin{align}
\label{eq:Q-box-linear-combi-bold}
\widehat{Q}_m({\bm \varepsilon}^{(d)},{\bm n}^{(d)})
 &=\sum_{\ell=1}^{d}\sum_{k=0}^{n_\ell-1}C_{\ell k}({\bm \varepsilon}^{(d)},{\bm n}^{(d)})
   \widehat{Q}_k({\varepsilon}_{\ell}),
\end{align}
where
\begin{align}
\label{eq:Ck-linear-combi-bold}
C_{\ell k}({\bm \varepsilon}^{(d)},{\bm n}^{(d)})
 &=\frac{1}
        {(n_{\ell}-k-1)!}
    \left(
     \frac{\partial}{\partial \varepsilon_{\ell}}
    \right)^{n_{\ell}-k-1}
    \left(
      \prod_{i=1(i\ne \ell)}^{d}
           \frac{1}
                {(\varepsilon_{l}-\varepsilon_{i})^{n_i}}
     \right).
\end{align}
The proof of the above expression is given in Appendix A. 
From the expansion formula for the multi-energy $\widehat{Q}_{m}$ box
we may conclude that the $E$-independent effective interaction $R$ for a 
system with non-degenerate unperturbed energies can be reduced to
calculating the $\widehat{Q}$ box and its energy derivatives with a single
energy variable. This situation is quite similar to the case with 
a degenerate unperturbed energy as has been shown in the former subsection.

   The convergence condition for the recursive solutions $\{ R_{n} \}$ in Eq.(\ref{eq:VII.13}) has
been discussed in Ref.\cite{SOEK94}. In the present case with non-degenerate unperturbed energies
the convergence condition is somewhat complicated because it depends on the initial states
$\{ |\phi_{\alpha}^{(0)}\rangle\}$ as well as the initial energies $\{ \varepsilon_{\alpha}\}$. However, it has been shown that, by appropriate choice of
$\{ |\phi_{\alpha}^{(0)}\rangle \}$ and $\{ \varepsilon_{\alpha} \}$, faster convergence is attained than in the
case with degenerate unperturbed energies~\cite{SOEK94}.
\section{Calculation of the $\widehat{Q}$ box and its derivatives}
\setcounter{equation}{0}
\subsection{A non-perturbative method for the $\widehat{Q}$ box}
In the present approach the $\widehat{Q}$ box is used as a building block of the
formulation. An essential problem in the derivation of the $E$-independent effective interaction
is how to calculate the $\widehat{Q}$ box and its energy derivatives, as has been shown in the former 
section. In place of the usual perturbative approach, we have proposed a non-perturbative
and recursive method for the $\widehat{Q}$ box in the previous study~\cite{SKMO13}.

We review briefly the method. We first introduce new subspaces in the $Q$ space. 
It has been shown that there exist subspaces $\{ Q_{1}, Q_{2}, \cdots\}$
in the $Q$ space such that they satisfy the conditions;
\begin{eqnarray}
\label{eq:Q-decomposition-conditions}
PHQ_{m}&=&Q_{m}HP=0 \ (m\ge 2),\\
Q_{m}HQ_{m+k}&=&Q_{m+k}HQ_{m}=0 \ (m\ge 1, \ k\ge 2).
\end{eqnarray}
The above conditions mean that the Hamiltonian is transformed to a block-tridiagonal form as 
\begin{eqnarray}
\label{eq:H-decomposition}
H=\begin{pmatrix}
    PHP      & PHQ_1     &      0        &    0        & \cdots \cr
    Q_1 HP  & Q_1 HQ_1 & Q_1 HQ_2  &    0         & \cdots \cr
    0          & Q_2 HQ_1 & Q_2 HQ_2  &  Q_2HQ_3 & \cdots \cr
    0          &       0      & Q_3 HQ_2  &  Q_3HQ_3 & \cdots \cr
    \vdots   & \vdots    & \vdots      & \vdots     &\vdots 
\end{pmatrix} 
.
\end{eqnarray}
The subspaces  $\{ P, Q_{1}, Q_{2},\cdots \}$ are essentially the same as the Krylov subspaces
\cite{GL96}.   The details of determining the subspaces  have been discussed in Ref.\cite{SKMO13}.

We consider a set of operators $\{\widetilde{e}_{1}(E), \,\widetilde{e}_{2}(E), \cdots, \widetilde{e}_n(E),\cdots\}$. The operator $\widetilde{e}_{n}(E)$ is a function of an energy variable $E$
and acts in the subspace $Q_n$. Suppose that  these operators obey the following descending 
recursion relation;
\begin{align}
\label{eq:tilde-e-(n-1)}
\widetilde{e}_n(E)
     &= {e}_n(E)-H_{n,n+1}\frac{1}{\widetilde{e}_{n+1}(E)}H_{n+1,n},
\end{align}
where 
\begin{align}
\label{eq:en_E}
e_{n}(E)&=Q_{n}(E-H)Q_{n},\\
\label{eq:H_nn+1}
H_{n,n+1}&=Q_{n}HQ_{n+1},
\end{align}
and
\begin{align}
\label{eq:H_n+1n}
H_{n+1,n}&=Q_{n+1}HQ_{n}.
\end{align}

We here assume that the system with the Hamiltonian $H$ can be well described 
in a finite dimensional Hilbert space.
In other words we assume that the $Q$ space can be truncated as 
\begin{align}
\label{eq:Q-decomposition}
 Q &= Q_1+Q_2+\cdots +Q_N
\end{align}
and the condition 
\begin{align}
\label{eq:convergence-condition}
 \left\|\frac{1}{{e}_{N+1}(E)}H_{N+1,N} \right\| \ll 1
\end{align}
is satisfied for a sufficiently large number $N$, 
where the symbol $\|X\|$ means the norm of a matrix $X$.

We then start the recursion in Eq.(\ref{eq:tilde-e-(n-1)}) with 
\begin{align}
\label{eq:tilde-e-(N)}
\widetilde{e}_{N}(E) &= e_N(E) \nonumber\\
                                &=Q_N (E-H) Q_N.
\end{align}
This recursion relation determines the sequence 
$\{\widetilde{e}_{N}(E), \,\widetilde{e}_{N-1}(E), \cdots, \widetilde{e}_1(E)\}$.
Recalling that the Hamiltonian $H$ maps the $P$-space states onto the $P$ space itself 
and only the $Q_1$ space, we may write as
\begin{align}
\label{eq:mapping-H}
 PHQ + \rm{h.c.} &= PHQ_1 + \rm{h.c.}.
\end{align}
Using the above relation, we finally have a simple expression of the $\widehat Q$ box as \cite{SKMO13}
\begin{align}
\label{eq:Q-box-tilde-e-(1)}
\widehat{Q}(E) &= PVP + PHQ_1\,\frac{1}{\widetilde{e}_{1}(E)}\,Q_1 HP.
\end{align}
The above formula means that the calculation of the $\widehat Q$ box can be reduced to 
that of the operator $\widetilde{e}_{1}(E)$. The $\widehat Q$ box can be 
expressed by only the second-order perturbation terms with 
the energy denominator $\widetilde{e}_{1}(E)$ and 
the vertices $PHQ_1+ \rm{h.c.}$.
%
\subsection{Calculation of the derivatives of the $\widehat{Q}$ box}
In order to calculate the $E$-independent effective interaction we need the energy
derivatives of the $\widehat{Q}$ box.
We here prove that these energy derivatives can be calculated analytically by means of 
a recursion method.

We introduce $\{ \widetilde{\xi}_n(E),\,\,n=1,2,\cdots, N \}$ as the inverse operators 
of $\{ \widetilde{e}_n(E),\,\,n=1,2,\cdots, N \}$, i.e.,
\begin{align}
\label{eq:definition-tilde-xi}
\widetilde{\xi}_{n}(E) &= \{\widetilde{e}_n(E)\}^{-1}
\end{align}
and write their energy derivatives as 
\begin{align}
\label{eq:energy-derivativ-tilde-en-(E)_0}
 \widetilde e^{(k)}_n(E) &= \frac{d^k}{dE^k} \widetilde e_n(E)
\end{align}
and
\begin{align}
\label{eq:energy-derivativ-tilde-en-(E)}
 \widetilde \xi^{(k)}_n(E) 
    &= \frac{d^k}{dE^k} \widetilde{\xi}_n(E)\cr
    &= \frac{d^k}{dE^k} \{\widetilde{e}_n(E)\}^{-1}.
\end{align}
We define for $k=0$ as
\begin{align}
\label{eq:definition-zeroth-tilde-en(E)_0}
\widetilde{e}^{(0)}_{n}(E) &= \widetilde{e}_n(E)
\end{align}
and
\begin{align}
\label{eq:definition-zeroth-tilde-en(E)}
\widetilde{\xi}^{(0)}_{n}(E) &= \widetilde{\xi}_n(E).
\end{align}

Using Eqs.(\ref{eq:Q-box-tilde-e-(1)}) and (\ref{eq:energy-derivativ-tilde-en-(E)}), 
$\widehat{Q}_k (E)$ in Eq.(\ref{eq:another-definition1-Q-box}) is written as
\begin{align}
\label{eq:another-definition2-Q-box}
 \widehat{Q}_k (E) &= \frac{1}{k!}PHQ_1\widetilde \xi^{(k)}_1(E)Q_1HP\,\,\,\,\,\,(k=0,1,2,\cdots).
\end{align}
For $k=0$, we define
\begin{align}
\label{eq:derivative-Q0-box}
 \widehat{Q}_0 (E) &= PHQ_1\widetilde \xi^{(0)}_1(E)Q_1HP\cr
                              &= PHQ_1\frac{1}{E-QHQ}Q_1HP.
\end{align}
The relation between the $\widehat{Q}$ box and $\widehat{Q}_0 (E)$ is
\begin{align}
\label{eq:relation-Q-box-Q0-box}
 \widehat{Q}(E) &= PVP + \widehat{Q}_0 (E).
\end{align}
The expression in Eq.(\ref{eq:another-definition2-Q-box}) means that the calculations of 
$\{\widehat{Q}_k (E) \}$ are reduced to those of $\{\widetilde{\xi}_1^{(k)}(E) \}$. 

   We show that $\{\widetilde{\xi}_1^{(k)}(E) \}$ can be given analytically 
through a recursion relation. We write the recursion relation in Eq.(\ref{eq:tilde-e-(n-1)}) 
for $\{\widetilde{e}_n(E) \}$  as
\begin{align}
\label{eq:tilde-e(n)-recursion}
\widetilde{e}_n(E)
     &= {e}_n(E)-H_{n,n+1}\widetilde{\xi}_{n+1}(E) H_{n+1,n},
\end{align}
where we have used Eq.(\ref{eq:definition-tilde-xi}). 
The $k$-times differentiations of the above equation lead to
\begin{align}
\label{eq:tilde-e(n)-k-th-recursion}
\widetilde{e}_n^{(k)}(E)
  &= \delta_{k0}e_n(E) +  \delta_{k1}Q_n - H_{n,n+1}\widetilde{\xi}_{n+1}^{(k)}(E) H_{n+1,n}.
\end{align}
For the calculations of $\{\widetilde{\xi}_n^{(k)}(E) \}$ we use the following Leibnitz formula : 
We start with the equality
\begin{align}
\label{eq:equality-tilde-e(n)-tilde-xi(n)}
\widetilde{e}_n(E) \cdot \widetilde{\xi}_n(E) = Q_n.
\end{align}
The $k$-times differentiations lead to 
\begin{align}
\label{eq:k-times-differentiations}
\sum_{m=0}^{k}\frac{k!}{m!(k-m)!}
         \widetilde{e}_n^{(m)}(E) \cdot \widetilde{\xi}_n^{(k-m)}(E) = 0,
\end{align}
from which we obtain
\begin{align}
\label{eq:k-times-differentiations-Leipnitz-formula}
\widetilde{\xi}_n^{(k)}(E) = -\sum_{m=1}^{k}\frac{k!}{m!(k-m)!}
        \widetilde{\xi}_n^{(0)}(E) \cdot \widetilde{e}_n^{(m)}(E) \cdot \widetilde{\xi}_n^{(k-m)}(E) .
\end{align}
We can prove that the operates $\{\widetilde{\xi}_n^{(k)}(E) \}$ can be calculated 
by the descending recursion relations in Eqs.(\ref{eq:tilde-e(n)-recursion}), (\ref{eq:tilde-e(n)-k-th-recursion}) 
and (\ref{eq:k-times-differentiations-Leipnitz-formula}). 
The starting values in the recursion relations are as follows : 
Let $N$ be the number of the subspaces $\{Q_m\}$ in the $Q$ space. 
We assume that for $n=N$ 
\begin{align}
\label{eq:tilde-e(N)-k-th-value}
\widetilde{e}_N^{(k)}(E)
  &= \left\{
       \begin{array}{ll}
          Q_N(E-H)Q_N  &\qquad k=0,\\
          Q_N              &\qquad k=1,\\
          0                   &\qquad k\geq 2
       \end{array}
       \right.
\end{align}
and
\begin{align}
\label{eq:tilde-xi(N)-0th}
\widetilde{\xi}_{N}^{(0)}(E) &= \{Q_N(E-H)Q_N\}^{-1}.
\end{align}

The calculation procedure for obtaining $\{ \widetilde{e}_m^{(k)}(E) \}$ and 
$\{ \widetilde{\xi}_m^{(k)}(E) \}$ is as follows : 
Suppose that they are given for 
$ n+1 \leq  m \leq N $ and $ 0 \leq  k \leq K $, 
where $K$ is the maximum number of the differentiations.  
For $m=n$ the operators 
$\{ \widetilde{e}_n^{(k)}(E), \,\,\, 0\le k \le K \}$ 
are given through the relation in Eq. (\ref{eq:tilde-e(n)-k-th-recursion}).
We next calculate
\begin{align}
\label{eq:tilde-xi(N)-0th-initial-value}
\widetilde{\xi}_{n}^{(0)}(E) &= \left\{ \widetilde{e}_n^{(0)}(E) \right\}^{-1},
\end{align}
which is the initial value in Eq.(\ref{eq:k-times-differentiations-Leipnitz-formula}). 
According to Eq.(\ref{eq:k-times-differentiations-Leipnitz-formula}) the operators 
$\{ \widetilde{\xi}_{n}^{(k)}(E), \,\,\, k=1,2,\cdots , K \}$ 
are calculated recursively. 
Repeating the manipulations we obtain all of the $\{ \widetilde{e}_m^{(k)}(E) \}$ and 
$\{ \widetilde{\xi}_m^{(k)}(E) \}$ which contain the operators 
$\{ \widetilde{\xi}_1^{(k)}(E)\}$ 
for $0 \leq k \leq K$. 
We then have all of the $\{ \widehat{Q}_k(E) \}$ for $ 0 \leq k \leq K $ as 
in Eq.(\ref{eq:another-definition2-Q-box}).

   As has been shown in the former section, the $E$-independent effective interaction can be 
expressed finally in terms of only the $\widehat{Q}$ boxes  
$\{ \widehat{Q}_{k}(E), k=1,2,\cdots\}$ 
with a single energy variable $E$. Therefore, the recursive solution 
 $\{ R_{n}, n=1,2,\cdots \}$ can be calculated analytically to arbitrary order for  
both systems with degenerate and non-degenerate unperturbed energies.
%
\section{MODEL CALCULATIONS}
\setcounter{equation}{0}
%
\subsection{The case with a degenerate unperturbed energy}
     In order to obtain some assessments of the present approach we study a
model problem.   
We start with a model Hamiltonian $H$ of which matrix elements are given by
\begin{eqnarray}
\label{eq:VIII.1}
\langle i|H|j\rangle &=&(\alpha i+\beta i^{2})\delta_{ij}+\gamma x_{ij}
\end{eqnarray}
with
\begin{eqnarray}
\label{eq:VIII.2}
 x_{ij} &=& 2\{ 
                    \sqrt{\sqrt{2}(i+j)} 
                  -[\sqrt{\sqrt{2}(i+j)}] 
                \}-1,
\end{eqnarray}
where $[x]$ is Gauss' notation which means the integer part of a real number $x$.
The $\alpha, \beta$, and $\gamma$ are the parameters chosen arbitrarily.   
A set of $\{ x_{ij}\}$ are recognized to be  pseudorandom numbers satisfying
\begin{eqnarray}
\label{eq:VIII.3}
-1 \le  x_{ij} \le 1. 
\end{eqnarray}
The model Hamiltonian is the same as that used in the previous study \cite{SKMO13}.
The total dimension of the matrix $H$  is taken to be  $N_{h}=100$.   As for the
model space (the $P$ space) we choose a two-dimensional space, {\it i.e.}, $d=2$.   
The basis states of the model space are taken to be two states which have the 
lowest and second-lowest diagonal matrix elements of  $H$.   With the 
Hamiltonian $H$  and the  $P$ space we introduce the subspaces (the Krylov 
subspaces) $\{ Q_{k}, k=1, 2, \dots\}$, where each subspace $Q_{k}$ is two-dimensional.
Since $N_{h}$ is the total dimension, the number $N_{q}$ of the Krylov subspaces is
given by
\begin{eqnarray}
\label{eq:VIII.4}
N_{q}&=&\frac{1}{2}(N_{h}-d)\nonumber\\
      &=&49.
\end{eqnarray}

   We consider a system with a degenerate unperturbed energy  $E_{0}$.   In this
model calculation we treat $E_{0}$ as an energy variable which is selected
arbitrarily.   We then modify the $P$-space part of the interaction as
\begin{eqnarray}
\label{eq:VIII.5}
PV'P&=&PHP-E_{0}P. 
\end{eqnarray}
The term $PVP$  in Eq.(\ref{eq:degenerate-Hamiltonian}) should be replaced with $PV'P$.   
This replacement ensures that $H$ has the same eigenvalues regardless of the
choice of the unperturbed energy $E_{0}$.

We first calculate the  $\widehat{Q}$ box for a given unperturbed energy by
using Eqs.(\ref{eq:tilde-e-(n-1)}) and (\ref{eq:Q-box-tilde-e-(1)}).  
We next calculate the energy-derivatives of
the $\widehat{Q}$  box and obtain  $\{ \widehat{Q}_{m}(E_{0}), m=1, 2, \cdots\}$ 
by following the calculation procedures given in 
Eqs.(\ref{eq:definition-tilde-xi})$\sim$(\ref{eq:tilde-xi(N)-0th-initial-value}).
     
With the $\widehat{Q}(E_{0})$  and   $\{ \widehat{Q}_{m}(E_{0}) \}$ we calculate the sequence
$\{ R_{1}, R_{2}, \cdots, R_{n}, \cdots\}$
by using the recurrence relation given in Eq.(\ref{eq:recursive-solution-Rn}).    
The eigenvalues of the $n$-th recursive solution $\{ E_{p}^{(n)}, p=1, 2, \cdots \}$ 
are obtained by diagonalizing the effective Hamiltonian as
\begin{eqnarray}
\label{eq:VIII.6}
H_{\rm eff}^{(n)} |\phi_{p}^{(n)}\rangle
&=& E_{p}^{(n)}|\phi_{p}^{(n)}\rangle,\,\,\,\,(p=1,2)
\end{eqnarray}
with
\begin{eqnarray}
\label{eq:VIII.7}
H_{\rm eff}^{(n)} &=& E_{0}P+R_{n}.
\end{eqnarray}
  The eigenvalue equation (\ref{eq:VIII.6}) determines two sequences 
$\{ E_{1}^{(n)}, n=1, 2, \cdots\}$  and  $\{ E_{2}^{(n)}, n=1, 2, \cdots\}$.       
  In Table \ref{table:convergence12} we show the calculated results of  $\{ E_{p}^{(n)}\}$ for
$p=1, 2$  and  $n=3, 6, 9, 12$   by taking the parameters to be $\alpha=2.0, \beta=0.4, \gamma=0.6$
 and $E_{0}=4.0$.
  We see that both of  $\{ E_{p}^{(n)}, p=1,2 \}$ converge to the exact eigenvalues 
$\{ E_{p},\,\,p=1,2\}$
  which are the lowest and second lowest eigenvalues of $H$.
  We have confirmed that the convergent eigenvalues are the nearest ones to the unperturbed energy 
$E_{0}=4.0$.
%
\begin{table}[!htb]
\caption
 {Convergence of the lowest two eigenvalues $E_{1}(n)$ and $E_{2}(n)$
 as functions of $n$, the number of recursions.   Here we employ the recursion
 method for a system with degenerate unperturbed energy which is chosen as
 $E_0 = 4.0$.  The parameters $\alpha,\beta$ and $\gamma$ are taken to be $\alpha=2.0$, $\beta=0.4$, and $\gamma=0.6$. 
Only the correct digits in each step of the recursion are presented. 
The exact values are $E_1 = 2.464968982871133$ and $E_2 = 5.468582694635212$. 
 }
  \label{table:convergence12}
\begin{center}
\begin{tabular}{lll} \hline \hline
  $n\,\,\,\,\,\,\,\,\,$     & $E_1(n)$        & $E_2(n)$       \\ \hline
  3       & 2.4649          & 5.468          \\
  6       & 2.46496         & 5.46858        \\
  9       & 2.46496898      & 5.468582       \\ 
12        & 2.464968982\,\,\,\,\,\,\,      & 5.46858269    \\ \hline\hline
\end{tabular}
 \end{center}
 \end{table}

\subsection{The case with non-degenerate unperturbed energies}
     We consider a system with the unperturbed Hamiltonian $H_{0}$ given in          
Eq.(\ref{eq:VII.2}).   
  In actual numerical calculations we treat the unperturbed 
energies $\{ \epsilon_{\alpha}\}$  and the corresponding unperturbed states 
$\{ |\phi_{p}^{(0)}\rangle\}$ as the input data which are selected arbitrarily.    
  In order to ensure that the effective interaction reproduces the same eigenvalues and eigenstates 
of the original Hamiltonian $H$, we modify the $P$-space interaction as
\begin{eqnarray}
\label{eq:VIII.10}
 PV'P&=&PHP-\sum_{\alpha}\epsilon_{\alpha}|\phi_{\alpha}^{(0)}\rangle
                                                   \langle{\tilde \phi}_{\alpha}^{(0)}|
\end{eqnarray}
and replace  $PVP$  in Eq.(\ref{eq:VII.1}) with  $PV'P$.

   We calculate the sequence  $\{ R_{1}, R_{2}, \cdots, R_{n}, \cdots\}$  by using 
the recursion relation in Eqs.(\ref{eq:VII.13})  and  (\ref{eq:VII.14}).    
  As has been proved in Subsection III.B, each of $\{ R_{n} \}$ can be calculated in terms of 
the $\widehat{Q}$ boxes $\{ \widehat{Q}(\epsilon_{\alpha}) \}$ in Eq.(\ref{eq:Q-box-tilde-e-(1)}) and 
$\{ \widehat{Q}_{m}(\epsilon_{\alpha}) \}$ which are the derivatives of the 
$\widehat{Q}$ boxes as in Eq.(\ref{eq:another-definition2-Q-box}).    
  The numerical results for  $\{ R_{n}\}$ and the corresponding 
eigenvalues  $\{ E_{p}^{(n)} \}$  depend on the choice of  $\{ \epsilon_{\alpha} \}$  and   
$\{ |\phi_{\alpha}^{(0)}\rangle \}$.    
  In Table \ref{table:convergence12ABCD}  we show the results for four cases with different sets of  
$\{ \epsilon_{\alpha} \}$  and  $\{ |\phi_{\alpha}^{(0)}\rangle \}$.    
  In this calculation the parameters  $\alpha, \beta$, and  $\gamma$ are chosen to be 
the same as those in Table.{\ref{table:convergence12}} .     
  The notations used are as follows :   
  The $|1\rangle$  and $|2\rangle$     are the original basis states which have the lowest and 
second-lowest diagonal matrix elements $\langle 1|H| 1\rangle$  and  
$\langle 2|H| 2\rangle$.   The $|\mu_{1}\rangle$  and  $|\mu_{2}\rangle$    
are the eigenstates of the $P$-space Hamiltonian $PHP$  written as
\begin{eqnarray}
\label{eq:VIII.11}
 PHP|\mu_{p}\rangle &=&E_{p}^{(0)}|\mu_{p}\rangle, \ (p= 1, 2).
\end{eqnarray}
\begin{table}[!htb]
\caption
{
  Convergence of the two lowest eigenvalues $E_1(n)$ and $E_2(n)$ as functions of $n$,
  the number of recursions, for four cases A, B, C and D with different unperturbed 
  energies $\varepsilon_1$ and $\varepsilon_2$ and initial states $|\phi_{1}^{(0)}\rangle$ and $|\phi_{2}^{(0)}\rangle$. \\
  A:  $\varepsilon_1$=2.0, $\varepsilon_2$=5.0, $|\phi_{1}^{(0)}\rangle=|1\rangle, |\phi_{2}^{(0)}\rangle=|2\rangle$,\ \,\,\,\,
  B:  $\varepsilon_1$=2.0, $\varepsilon_2$=5.0, $|\phi_{1}^{(0)}\rangle=|\mu_{1}\rangle, |\phi_{2}^{(0)}\rangle=|\mu_{2}\rangle$,\\
  C:  $\varepsilon_1$=2.5, $\varepsilon_2$=5.5, $|\phi_{1}^{(0)}\rangle=|1\rangle, |\phi_{2}^{(0)}\rangle=|2\rangle$,\ \,\,\,\,
  D:  $\varepsilon_1$=2.5, $\varepsilon_2$=5.5, $|\phi_{1}^{(0)}\rangle=|\mu_{1}\rangle, |\phi_{2}^{(0)}\rangle=|\mu_{2}\rangle$.\\
The definitions of $ |1\rangle,|2\rangle,|\mu_{1}\rangle$ and $ |\mu_{2}\rangle$ are 
given in the text. 
The exact values for $E_1$ and $E_2$ are given in Table \ref{table:convergence12}.
}
  \label{table:convergence12ABCD}
\begin{center}
\begin{tabular}{llllll} \hline \hline
      \,\,\,\,\,\,\,\,\,\,\,\,\,\,\,& $n$\,\,\,\,\,\,\,\,\,\, &  A         & B             & C            & D                 \\ \hline
      & 2    & 2.46       & 2.464         & 2.46         & 2.46496           \\
 $E_1$& 4    & 2.46496    &  2.46496      & 2.46496      & 2.464968982871    \\
      & 6    & 2.46496\,\,\,\,\,\,\,\,     &  2.464968982\,\,\,\,\,\,\,\,   & 2.464968982\,\,\,\,\,\,\,\,   & 2.464968982871133 \\ \hline           
         & 2          & 5.468        &  5.468             & 5.468         & 5.46858\\ 
 $E_2$   & 4          & 5.46858      &  5.46858           & 5.46858       & 5.4685826946\\  
         & 6          & 5.468582     &  5.468582          & 5.468582      & 5.46858269463521\\ \hline\hline
\end{tabular}
 \end{center}
 \end{table}
  We employ two sets of  $\{ |1\rangle, |2\rangle \}$  and $\{|\mu_{1}\rangle, |\mu_{2}\rangle \}$ 
as the unperturbed states $\{ |\phi_{1}^{(0)}\rangle, |\phi_{2}^{(0)}\rangle \}$.    
  As for the unperturbed energies  
$\{ \epsilon_{1}, \epsilon_{2} \}$   we choose two sets of (2.0, 5.0) and (2.5, 5.5). 
 In Table \ref{table:convergence12ABCD} the results are given for four combinations of  
$\{ \epsilon_{1}, \epsilon_{2} \}$  and  $\{ |\phi_{1}^{(0)}\rangle, |\phi_{2}^{(0)}\rangle \}$.   
  In all of the cases the 
convergent solutions for the lowest and second-lowest eigenvalues of $H$,
namely,  $E_{1}$  and $E_{2}$      are reproduced.     
  It is clear that the fastest 
convergence is attained for the case D where the unperturbed energies and states are taken to be  
$(\epsilon_{1}, \epsilon_{2})=(2.5, 2.5)$  and
$\{ |\phi_{1}^{(0)}\rangle,|\phi_{2}^{(0)}\rangle\} =\{ |\mu_{1}\rangle,|\mu_{2}\rangle \}$.  
  Comparing the present results with those in Table  \ref{table:convergence12}, 
we see that the convergence is much faster than that in the case with a degenerate unperturbed energy.
   We may conclude that, even though the calculation procedure in the case
with non-degenerate unperturbed energies is rather complicated, it has some 
advantages in bringing about faster convergence than the case with a
degenerate unperturbed energy.
%
\section{Concluding Remarks}
     The present status of the effective-interaction theory for a few
valence particles in nuclei would be summarized as follows:  
  Various methods have been given, which include the KK iteration method~\cite{KK74},
the $\widehat{Z}$ box method~\cite{SOKF11},  the LS recursion method~\cite{LS80, SL80}, 
the generalized LS (GLS) method~\cite{SOEK94} and the extended KK method of Takayanagi
\cite{Tak11}.    
  All these solutions for the effective interaction given to date are represented in
terms of the $\widehat{Q}$ box and its derivatives introduced by Kuo and others~\cite{KLR71}.
  The $\widehat{Q}$ box itself has been calculated usually in a perturbative way.   
Some of the numerical calculations seem to confirm that dominant contribution 
comes from second-order terms and the other higher-order terms are less 
important. 
  The LS scheme derives an $E$-independent solution, but we
need to calculate the derivatives of the $\widehat{Q}$ box of higher order if we want 
to perform calculations of higher order.   
  The derivatives of the $\widehat{Q}$ box have been calculated by using the method 
of numerical differentiation.
  The calculation of higher-order derivatives requires the 
$\widehat{Q}$ boxes at many points of the energy variable $E$.   
  The numerical differentiation of arbitrary order is not an easy task to complete.

    For these reasons a remaining problem for further development of the
effective interaction theory has been to find out a method for calculating the $\widehat{Q}$ box
and its derivatives more simply and accurately.   
  The non-perturbative recursion method has been given for the $\widehat{Q}$ box 
and its derivatives up to second order in the previous work by the authors~\cite{SKMO13}.  
 In the present study we have extended the
recursion method to calculate the derivatives of arbitrary order.   
  With these recursive solutions we have shown that the LS and the GLS solutions for the
effective interaction can be calculated up to arbitrary order.   
  This method is characterized to be $E$-independent and non-perturbative.  
It will be a quite interesting problem to compare the present non-perturbative solution with that in 
the usual perturbation method.  
We believe that the present non-perturbative approach would
mark one step toward developing the effective-interaction theory.
%
%

%
%
\appendix
\section{}
The multi-energy $\widehat{Q}$ box defined in Eq.(\ref{eq:Q-box-eps-n}) can be
expanded into a linear combination of $\{ \widehat{Q}_{k}(\varepsilon_{i}), k=1,2,\cdots\}$
with a single energy variable $\varepsilon_{i}$. We here prove that the coefficients 
$\{ C_{\ell k}({\bm \varepsilon}^{(d)},{\bm n}^{(d)}) \}$ in Eq.(\ref{eq:Q-box-linear-combi-bold})
can be given as in Eq.(\ref{eq:Ck-linear-combi-bold}).

We first note the equality  
\begin{align}
\label{eq:fraction-n}
\frac{1}{(E-QHQ)^{n}}
 &=\frac{(-1)^{n-1}}{(n-1)!}
   \frac{d^{n-1}}{dE^{n-1}} \left( 
                             \frac{1}{E-QHQ}
                            \right),                        
\end{align}
which follows that the multi-energy $\widehat{Q}_{m}$ box  in Eq.(\ref{eq:Q-box-eps-n})
can be given by
\begin{align}
\label{eq:Q-m-bold-1}
 \widehat{Q}_{m}({\bm \varepsilon}^{(d)},{\bm n}^{(d)})
 &=(-1)^{m}PHQ
   \left[ 
     \prod_{i=1}^{d}  
    \frac{(-1)^{n_i-1}}
        {(n_{i}-1)!}
    \left(
     \frac{\partial}{\partial \varepsilon_{i}}
    \right)^{n_{i}-1}
    \left(
      \prod_{j=1}^{d}
           \frac{1}
                {(\varepsilon_{j}-QHQ)}
     \right)
   \right]QHP\nonumber\\
&=\prod_{i=1}^{d}  
    \frac{1}
         {(n_{i}-1)!}
    \left(
     \frac{\partial}
          {\partial \varepsilon_{i}}
    \right)^{n_i-1}
    \widehat{Q}_{d-1}(\varepsilon_1,\varepsilon_2,\cdots,\varepsilon_d).
\end{align}
Substituting Eq.(\ref{eq:Q-box-linear-combi}) into Eq.(\ref{eq:Q-m-bold-1}), 
we may write the $\widehat{Q}_{m}$ box in terms of the 
$\widehat{Q}$-boxes $\{\widehat{Q}(\varepsilon_i)\}$ as
\begin{align}
\label{eq:Q-m-bold-2_0}
 \widehat{Q}_{m}({\bm \varepsilon}^{(d)},{\bm n}^{(d)})
 = \sum_{l=1}^d \prod_{i=1}^{d} 
    \frac{1}{(n_{i}-1)!}
    \left(
     \frac{\partial}
          {\partial \varepsilon_{i}}
    \right)^{n_i -1}
    C_l(\varepsilon_1,\varepsilon_2,\cdots,\varepsilon_d)\widehat{Q}(\varepsilon_\ell).
\end{align}
We note that, from the definition of $C_l(\varepsilon_1,\varepsilon_2,\cdots,\varepsilon_d)$ 
in Eq.(\ref{eq:Ck-linear-combi}), the following equality can be derived;
\begin{align}
\label{eq:equality}
\prod_{i=1(i \neq l)}^{d} 
    \frac{1}{(n_{i}-1)!}
    \left(
     \frac{\partial}
          {\partial \varepsilon_{i}}
    \right)^{n_i -1}
    C_l(\varepsilon_1,\varepsilon_2,\cdots,\varepsilon_d)
    =\prod_{i=1(i \neq l)}^{d} \frac{1}{(\varepsilon_l -\varepsilon_i)^{n_i}} .
\end{align}
Using the above relation we can write the $\widehat{Q}_{m}$ box in the form 
\begin{align}
\label{eq:Q-m-bold-3}
 \widehat{Q}_{m}({\bm \varepsilon}^{(d)},{\bm n}^{(d)})
 = \sum_{l=1}^d 
    \frac{1}{(n_{l}-1)!}
    \left(
     \frac{\partial}
          {\partial \varepsilon_{l}}\right)^{n_l -1}
    \left[\prod_{i=1(i \neq l)}^{d} \frac{1}{(\varepsilon_l -\varepsilon_i)^{n_i}}\right]\widehat{Q}{(\epsilon_l)}.
\end{align}
Applying the Leibnitz formula for the differentiation of a product of two functions, we have 
\begin{align}
\label{eq:Q-m-bold-4}
 \widehat{Q}_{m}({\bm \varepsilon}^{(d)},{\bm n}^{(d)})
 = \sum_{l=1}^d \sum_{k=0}^{n_l -1}G_{lk}
     \left[\frac{1}{k!}
     \left(
     \frac{\partial}
          {\partial \varepsilon_{l}}\right)^{k}
   \widehat{Q}{(\epsilon_l)}\right] 
\end{align}
with 
\begin{align}
\label{eq:G-lk}
 G_{lk}=\frac{1}{(n_l -k-1)!}
     \left(
     \frac{\partial}
          {\partial \varepsilon_{l}}\right)^{n_l - k-1}
   \left[\prod_{i=1(i \neq l)}^{d} \frac{1}{(\varepsilon_l -\varepsilon_i)^{n_i}}\right].
\end{align}
Making use of the definition of  $\widehat{Q}_k{(\epsilon_l)}$ in Eq.(\ref{eq:another-definition1-Q-box}), 
we finally obtain the expression of the $\widehat{Q}_m$ box as
\begin{align}
\label{eq:Q-m-bold-2}
 \widehat{Q}_{m}({\bm \varepsilon}^{(d)},{\bm n}^{(d)})
 &=\sum_{\ell=1}^{d}\sum_{k=0}^{n_\ell-1}
    \left[
     \frac{1}{(n_\ell -k-1)!}
      \left(
       \frac{\partial}
            {\partial \varepsilon_{l}}
      \right)^{n_\ell-k-1}
      \left(
      \prod_{i=1(i\ne \ell)}^{d}
           \frac{1}
                {(\varepsilon_{\ell}-\varepsilon_{i})^{n_i}}
     \right)
    \right]
           \widehat{Q}_{k}(\varepsilon_\ell).
\end{align}
The above formula implies  that the coefficient $C_{\ell k}({\bm \varepsilon}^{(d)},{\bm n}^{(d)})$
is given as in Eq.(\ref{eq:Ck-linear-combi-bold}) and the multi-energy $\widehat{Q}_{m}$ box 
can be expanded as in Eq.(\ref{eq:Q-box-linear-combi-bold}).

We show a simple example of the multi-energy $\widehat{Q}$ box for the case of the two-dimensional
$P$ space $(d=2)$. 
Let $\varepsilon_1$ and $\varepsilon_2$ be the unperturbed energies.
 We write the $\widehat{Q}_{m}$ box as
\begin{align}
\label{eq:Qm-2dim-1}
 \widehat{Q}_{m}(\varepsilon_1,\varepsilon_2,n_1,n_2)
  &=(-1)^{m}PHQ\frac{1}
                    {(\varepsilon_1-QHQ)^{n_1}(\varepsilon_2-QHQ)^{n_2}}
            QHP\delta_{m,n_1+n_2-1}.
\end{align}
Using Eq.(\ref{eq:G-lk}), we have 
\begin{align}
\label{eq:Qm-2dim-2}
 \widehat{Q}_{m}(\varepsilon_1,\varepsilon_2,n_1,n_2)
  &=\sum_{\ell=1}^{2}\sum_{k=0}^{n_\ell-1}
     C_{\ell k}(\varepsilon_1,\varepsilon_2,n_1,n_2)
     \widehat{Q}_{k}(\varepsilon_\ell)
\end{align}
with
\begin{align}
\label{eq:Cell-k-2dim}
 C_{\ell k}(\varepsilon_1,\varepsilon_2,n_1,n_2)
  &=\delta_{m,n_1+n_2-1}
     \frac{(-1)^{n_\ell+k+1}(m-k-1)!}
          {(m-n_\ell)!(n_\ell-k-1)!(\varepsilon_\ell-\varepsilon'_{\ell})^{m-k}}     
     \widehat{Q}_{k}(\varepsilon_\ell),
\end{align}
where $\varepsilon'_{\ell}=\varepsilon_2$ for $\ell=1$ and
      $\varepsilon'_{\ell}=\varepsilon_1$ for $\ell=2$.

We finally note that any of the $\widehat{Q}$ boxes, including the multi-energy $\widehat{Q}_{m}$ boxes,
 is an operator acting in the $P$ space. Therefore it has a $d\times d$ matrix representation, where
 $d$ is the dimension of the $P$ space. Usually the dimension $d$ is taken to be a small number.
 If the $\widehat{Q}$ boxes with a single energy variable are given analytically,
 the multi-energy $\widehat{Q}_{m}$ boxes can also be calculated exactly without any approximation,
 because the r.h.s. of Eq.(\ref{eq:Q-m-bold-2}) is merely a linear combination of the $d\times d$
 matrices $\{ \widehat{Q}_{k}(\varepsilon_\ell), k=0,1,\cdots\}$.

%
 

\end{document}